\documentclass[8pt]{article}  \usepackage{times}
\usepackage{graphicx}

\usepackage{color}

\usepackage[round,numbers,sort&compress]{natbib} 

\usepackage{amsmath}
\usepackage{amssymb}
\usepackage{stackrel}
\usepackage{chemarrow}
\usepackage{graphicx}

\usepackage{booktabs}
\usepackage{dcolumn}
\usepackage{rotating}
\usepackage{multirow}

\usepackage[compact]{titlesec}

\begin{document}

\titlespacing{\section}{0pt}{2pt}{0pt}
\twocolumn[{\LARGE \textbf{The thermodynamics of general and local anesthesia\\*[0.2cm]}}
{\large Kaare Gr\ae sb\o ll, Henrike Sasse-Middelhoff and Thomas Heimburg$^{\ast}$\\*[0.1cm]
{\small Niels Bohr Institute, University of Copenhagen, Blegdamsvej 17, 2100 Copenhagen \O, Denmark}\\

{\normalsize ABSTRACT\hspace{0.5cm} General anesthetics are known to cause depression of the freezing point of transitions in biomembranes. This is a consequence of ideal mixing of the anesthetic drugs in the membrane fluid phase and exclusion from the solid phase. Such a generic law provides physical justification of the famous Meyer-Overton rule. We show here that general anesthetics, barbiturates and local anesthetics all display the same effect on melting transitions. Their effect is reversed by hydrostatic pressure. Thus, the thermodynamic behaviour of local anesthetics is very similar to that of general anesthetics. We present a detailed thermodynamic analysis of heat capacity profiles of membranes in the presence of anesthetics. This analysis is able to describe experimentally observed calorimetric profiles and permits prediction of the anesthetic features of arbitrary molecules. In addition, we discuss the thermodynamic origin of the cutoff-effect of long-chain alcohols and the additivity of the effect of general and local anesthetics. 
\\*[0.0cm] }}
]

\noindent\footnotesize {$^{\ast}$corresponding author, theimbu@nbi.dk. Present affiliation of K.G. is the Technical University of Denmark.}\\

\noindent\footnotesize{\textbf{Abbreviations:} DPPC - 1,2-dipalmitoyl-sn-glycero-3-phosphatidylcholine; LUV - large unilamellar vesicles ({\O} $\sim$ 100 nm); DSC - differential scanning calorimetry}\\

\normalsize
\section*{Introduction}

General anesthetics (including barbiturates) and local anesthetics were introduced to clinical praxis during the mid 19th century. The first general anesthetic used in surgery was diethyl ether, but many others have been found.  These include nitrous oxide (laughing gas), propofol, halothane, sevoflurane \cite{Urban2002} but also the chemically inert noble gas xenon and barbiturates \cite{Lopez-Munoz2005} --- molecules that are very different structurally.  The first local anesthetic was cocaine, whose analgesic effect was described in the late 1850s.  Most known local anesthetics were later developed in an attempt to avoid the addictive effect of cocaine \cite{Ruetsch2001}. Thus, the distinction between the different classes of anesthetics is partially of historical origin. Today, it is generally believed that general and local anesthetics act by different mechanisms. Local anesthesia has frequently been attributed to the specific interaction of local anesthetics with (sodium) channel proteins \cite{Butterworth1990, Fozzard2011}. Simultaneously, there is wide agreement that the action of general anesthetics is not well understood.  However, there exists a striking inverse linear correlation between the solubility of general anesthetics in the lipid membrane and the critical anesthetic dose known as the Meyer-Overton correlation \cite{Overton1901, Overton1991, Urban2002}.  It applies to drugs of quite different chemical structure such as nitrous oxide (laughing gas), xenon (a noble gas) and sevoflurane (a fluorinated organic solvent).   Another way of stating the Meyer-Overton correlation is $[ED_{50}]\cdot P$=const., where $[ED_{50}]$ is the effective anesthetics concentration in the alveolar volume or in the blood where 50\% of individuals are anesthetized, and $P$ is the partition coefficient between oil and water (or air).  At critical dose the membrane concentration of all general anesthetics is exactly the same \cite{Heimburg2007c}. The Meyer-Overton correlation leads to the notion that general anesthesia is closely related to the solubility of the drug in the lipid membrane.  Indeed, in his book from 1901  \cite{Overton1901} Overton proposed that the correlation between membrane partitioning and critical dose suggests the existence of a generic physical mechanism for anesthesia.   However, Overton did not provide any such mechanism, and his correlation remained a true but unexplained observation. 

In the absence of such an explanation, some researchers favor a view involving binding to molecular targets. In particular Franks and collaborators made this case popular using firefly luciferase as a target protein\cite{Moss1991, Franks1994, Franks2008}. They showed that the equilibrium between two structural forms of the fluorescing luciferase is controlled by most (but not all) general anesthetics. Similar mechanisms have been thought to apply for membrane proteins. This view is probably incompatible with a unique mechanism for anesthesia.  As argued above, at critical dose the membrane concentration of all general anesthetics is identical --- including the noble gas xenon. Thus, if both the Meyer-Overton correlation and binding to a molecular target are true, the equilibrium association constant between membrane-dissolved drug and membrane protein must be exactly the same value for all drugs including xenon, which is an inert noble gas. 

Cantor \cite{Cantor1997a, Cantor1997b} proposed to combine an unspecific lipid solubility with a protein mechanism by investigating the lateral pressure profile of membranes in the presence of anesthetics.  Anesthetics behave like ideal gases that exert forces on interfaces, and this pressure putatively alters protein structure and function. Since lateral pressure profiles would be affected in a very similar manner by different general anesthetics, it is suggested that anesthetics would alter a protein structure in a generic manner. Unfortunately, Cantor's view remains a speculation as long as the precise structures and functions of individual protein conformations are unknown. However, Cantor and collaborators provide some experimental evidence for currents induced in GABA$_A$ receptors by isoflurane and sevoflurane \cite{Cantor2009}.

It seems most likely that the influence of anesthetics on the thermodynamics of membranes contains the key for explaining the above correlation. We and other authors observed that general anesthetics induce a lowering of the solid-liquid transition temperature in lipid membranes (e.g., \cite{Trudell1975, Kaminoh1992, Kharakoz2001, Heimburg2007c}). The melting point of membranes has exactly the same correlation with the critical anesthetic dose as the partition coefficient. We showed that this effect can be well explained by the so-called freezing point depression law that originates from van\,'t Hoff.   When applied to membranes, this law involves a slight modification of the Meyer-Overton relation.  Thus, we proposed that the anesthetics are ideally soluble only in the liquid phase of the membrane but are insoluble in the solid phase. The shift of the transition is now due to the difference in the entropy of mixing in the two phases. This explanation of anesthesia has the virtue of explaining several properties of general anesthetics:
\begin{itemize}
  \item Excess hydrostatic pressure increases melting temperatures \cite{Trudell1975, Mountcastle1978, Ebel2001} and opposes the effect of anesthetics on transitions. This can explain the effect of the pressure-reversal of anesthesia quantitatively \cite{Johnson1950, Halsey1975, Heimburg2007c}. Anesthetic inhibition of the protein firefly luciferase, in contrast, does not display pressure reversal \cite{Moss1991}.
  \item The freezing-point depression law is based on the notion of ideal solutions in which individual molecules do not interact.   The shift is thus linear in the concentration of anesthetics in the membrane. Therefore, freezing-point depression is consistent with the observed additivity of the anesthetic action \cite{Overton1901, diFazio1972, Fang1997}.
  \item Many membrane-soluble molecules do not dissolve ideally in liquid membranes, or they dissolve in the solid phase as well. Such molecules would not act as general anesthetics. For instance, cholesterol does not lower (but rather increases) melting temperatures \cite{Mabrey1978} and therefore does not display anesthetic activity.
  \item Molecules that display phase behavior of their own should not behave as ideal anesthetics. This includes most lipids or the long-chain alcohols (starting from chain lengths of twelve, i.e., dodecanol) that display melting points above room temperature and do not act as anesthetics \cite{Pringle1981}. This effect is known as the cutoff-effect (see discussion).
\end{itemize}
Freezing point depression by anesthetics is also consistent with a recent theory for nerve pulse propagation that is based on phase transitions in biological membranes  \cite{Heimburg2005c, Heimburg2007b, Andersen2009}. Close to transitions in membranes, solitary electromechanical waves are possible. Such melting transitions in fact have been found for a number of biomembranes \cite{Heimburg2007a}. 
Anesthetics render the excitation of the pulse more difficult while hydrostatic pressure facilitates it \cite{Heimburg2007c, Heimburg2008}. 

No law similar to the Meyer-Overton correlation exists for local anesthetics.  At high concentrations these anesthetics are toxic \cite{Cox2003}. Local anesthetics are typically not administered intravenously, and a critical dose cannot be determined. This has generally lead to the notion that local anesthetics work by a different mechanism, e.g., by blocking sodium channels \cite{Butterworth1990, Fozzard2011}. However, this is not necessarily inconsistent with local anesthetics having general anesthetic properties. The Meyer-Overton correlation implies that the effect of general anesthetics is additive.  Two different anesthetics each with half-critical concentration yield full anesthesia. Interestingly, it has been reported by various authors that the effects of local and general anesthetics are additive (e.g., \cite{Himes1977, Ben-Shlomo2003}, see discussion section). This finding is striking since it is difficult to reconcile with the assumption that local and general anesthesia work by different mechanisms. One rather would conclude that local anesthetics display general anesthetic properties. In this context it interesting to note that local anesthetics are also membrane active and lower melting transitions \cite{Papahadjopoulos1975, Winter1991, Ueda1994, Matsuki1996, Hata2000, Lygre2009, Paiva2012}. It is therefore plausible to hypothesize that local anesthetics have properties of general anesthetics.

In this publication we investigate the melting behavior of membranes in the presence of both general and local anesthetics. We provide a thermodynamic formalism to understand the heat capacity profiles of lipid membranes containing both general and local anesthetics. We show that the same theoretical treatment applies to both classes of anesthetics and that their effects on transition temperatures is equally reversed by hydrostatic pressure.


\section*{Materials and Methods}\label{Methods}
{\small Lipids were purchased from Avanti Polar Lipids (Birmingham, AL) and used without further purification. Octanol was purchased from Fluka (Buchs, Switzerland). All other anesthetics were purchased from Sigma-Aldrich (St. Louis, MO). Multilamellar lipid dispersions (10mM, buffer: 10mM Hepes, 1mM EDTA, pH 7.0, octanol concentration adjusted) were prepared by vortexing the lipid dispersions above the phase transition temperature of the lipid.

To generate multilamellar vesicles, dry lipids were dissolved in the buffer and vortexed above the main phase transition temperature until no visible clumps remained. For the generation of unilamellar vesicles (LUV), multilamellar vesicles were extruded at least 30 times above the phase transition temperature of the respective lipid using an extruder (Avestin Europe, Mannheim, Germany) and filters with a pore size of 100nm. The resulting large unilamellar vesicles (LUV) are stable in the refrigerator for at least two weeks.

Anesthetics were added to DPPC in two distinct ways. One was simply to dissolve the anesthetic in the buffer and then follow the recipe given above. This method has the advantage that the small concentrations needed could be obtained easily by dissolving in large amounts of buffer and then further diluting with the addition of pure buffer. The disadvantage of this method is that many of the anesthetics are so hydrophobic that even small amounts are impossible to dissolve. Under these conditions, we dissolved anesthetics in a 2:1 methanol-dichloromethane mixture and added the solution to the dry lipids in appropriate quantities. The solution was dried under an air stream and subsequently dried further over night in a high vacuum desiccator to remove any remaining solvent. The dry lipid films containing anesthetics were than hydrated with buffer and extruded as described above. This method only work for non-volatile anesthetics.

Heat capacity profiles were obtained using a VP-scanning calorimeter (MicroCal, Northampton, MA) at scan rates of 5 deg/hr. The curves presented in this work are down-scans from high to low temperature. Since anesthetics are soluble in the fluid phase it seems plausible to assume that the samples equilibrate faster during down-scans. Due to the extrusion process, the total amount of lipid in the dispersion obtained after extrusion may vary slightly. All experimental profiles were renormalized to a constant transition enthalpy of 35 kJ/mol. The shapes of the heat capacity profiles of extruded vesicles also can also vary slightly in time. This may explain some of the minor variations in shape observed with different concentrations of anesthetics. A typical experiment lasted several days. Repeating an experiment with new samples under similar conditions yielded similar but not necessarily exactly identical profiles. 
}

\section*{Theory}\label{Theory}
\begin{figure}[htb!]
    \centering
	\includegraphics[width=8cm]{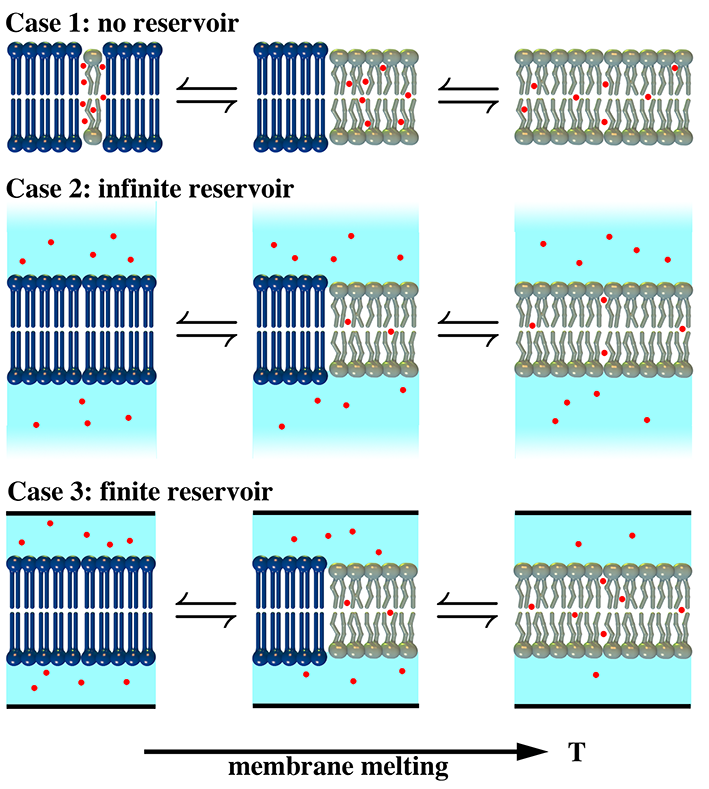}
	\parbox[c]{8.5cm}{ \caption{\textit{Three different scenarios of a membrane in presence of anesthetics. Case 1: The aqueous volume is very small or the partition coefficient of the anesthetic in the fluid membrane is very high. All anesthetic molecules reside in the fluid lipid phase independent of its volume. If the fraction of fluid phase changes as a function of temperature, the concentration of anesthetics in the fluid membrane is also changed.  Case 2: The aqueous volume is infinitely large.  The concentration of anesthetics in both the aqueous phase and in the fluid membrane is constant and independent of temperature. Case 3: The aqueous volume is finite. When the amount of fluid phase changes, the concentration of anesthetics changes in both the aqueous and the fluid membrane . This is the general case.}
	\label{Figure0}}}
\end{figure}

In the following we will present the thermodynamic analysis of a chemically inert drug ideally soluble in the fluid lipid membrane but insoluble in the gel membrane. This scenario was shown to describe the thermodynamic behavior of general anesthetics in biomembranes \cite{Heimburg2007c}.
In the following we will discuss three cases:
\begin{enumerate}
  \item The absence of an aqueous reservoir, or equivalently, drug insolubility in the buffer (i.e., an infinite partition coefficient in the fluid membrane). All anesthetic drugs reside in the membrane. The melting of the membrane alters the concentration of the anesthetics in the fluid membrane because the fraction of fluid phase changes.
  \item Infinite aqueous reservoir with constant anesthetics concentration. In this case, the concentration of anesthetics in the fluid membrane is constant.The ratio of the two concentrations is given by the partition coefficient.
  \item Finite aqueous reservoir. During melting, both the concentrations of anesthetics in the buffer and in the fluid membrane change. 
\end{enumerate}
These case are schematically depicted in Fig.\,\ref{Figure0}. The goal is to provide a general method for the interpretation of heat capacity profiles of membranes in an aqueous buffer containing anesthetic drugs.

\subsection*{1. Highly membrane-soluble anesthetics}\label{Theory_1}
Consider the mixture of a lipid membrane with melting temperature $T_m$, and an anesthetic molecule, $A$, that dissolves ideally in the fluid phase but not in the gel phase of the lipid. We do not consider an aqueous phase, i.e., we assume that practically all drugs dissolve in the membrane and cannot dissociate from the membrane. This case was considered by us previously \cite{Heimburg2007c}. The phase diagram of such a mixture is given by ideal solution theory.  The chemical potentials of the fluid and the gel membrane are given
\begin{eqnarray}
\label{T1.1}
\mu_L^f&=&\mu_{L,0}^f+RT \ln x_L^f\nonumber\\
\mu_L^g&=&\mu_{L,0}^g \;,
\end{eqnarray}
where $x_L^f$ is the molar fraction of lipid and $x_A^f=1-x_L^f$ is the molar fraction of anesthetics in the fluid phase. If the total membrane is in the fluid state, the total fraction of anesthetics, $x_A$, is identical to $x_A^f$. In equilibrium, $\mu_L^f=\mu_L^g$, and therefore
\begin{equation}
\label{T1.2}
\ln x_L^f=-\frac{\mu_{L,0}^f-\mu_{L,0}^g}{RT} \;.
\end{equation}
The difference of the standard chemical potentials, $\Delta \mu_{L,0}=\mu_L^f-\mu_{L,0}^g$, is given by $\Delta H_{L,0}-T\Delta S_{L,0}$, where $\Delta H_{L,0}$ is the molar enthalpy of melting and $\Delta S_{L,0}$ is the molar entropy of melting. We arrive at
\begin{equation}
\label{T1.3}
\ln x_L^f=-\frac{\Delta H_{L,0}-T\Delta S_{L,0}}{RT} \;.
\end{equation}
The melting temperature in the absence of anesthetics is given by $T_{m,L}=\Delta H_{L,0}/\Delta S_{L,0}$. Thus,
\begin{equation}
\label{T1.4}
\ln x_{L,0}^f=-\frac{\Delta H_{L,0}}{R}\left(\frac{1}{T}-\frac{1}{T_{m,L}}\right)\equiv \ln(1-x_A^f) \;.
\end{equation}
By approximating $\ln(1-x_A^f)\approx -x_A^f$ and $T\cdot T_{m,L}\approx T_{m,L}^2$, we arrive at
\begin{equation}
\label{T1.5}
\Delta T = T-T_{m,L}=-\frac{R T_{m,L}^2}{\Delta H_{L,0}}x_A^f \;.
\end{equation}
This is the well known freezing-point depression law originating from J. H. van\,'t Hoff \cite{vantHoff1886}.

\begin{figure}[htb!]
    \centering
	\includegraphics[width=8cm]{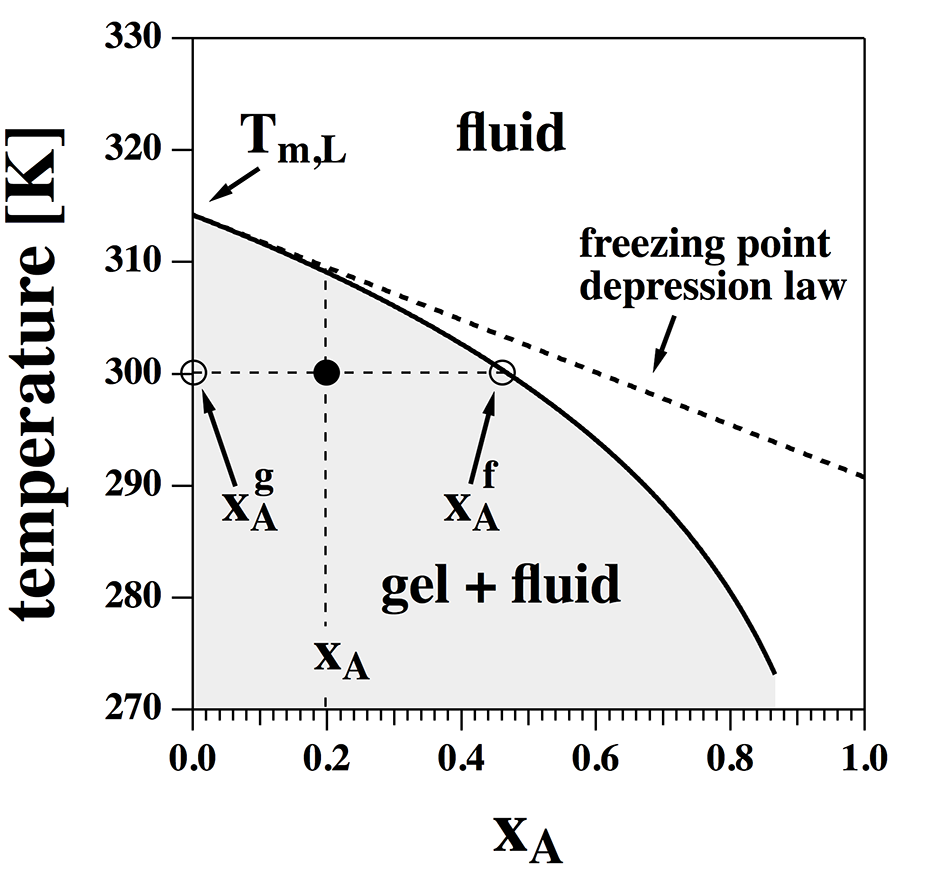}
	\parbox[c]{8.5cm}{ \caption{\textit{The phase diagram for case 1 in Fig.\,\ref{Figure0} for DPPC vesicles in the presence of a general anesthetic. The solid line indicates the concentration of anesthetics in the fluid phase as a function of temperature (fluids line). The concentration in the gel phase is zero by definition. The solid line also indicates the onset of the lipid melting transition upon cooling at a given anesthetic concentration. The dotted line indicates the freezing point depression approximation given by eq. (\ref{T1.5}). It is valid up to about 20 mol\% of anesthetics and deviates strongly at higher concentrations. For $x_A=0.2$ (total membrane fraction of anesthetics) and T=300 K, the fluid and solid fractions are $x_A^f=0.47$ and $x_A^g=0$. }
	\label{Figure1}}}
\end{figure}
The laws given in eqs.\,(\ref{T1.4}) and (\ref{T1.5}) are represented graphically in Fig.\,\ref{Figure1}. We chose the parameters for DPPC membranes, i.e., $\Delta H_{L,0}=35 \,{\rm kJ/mol}$ and $T_{m,L}=314.2\, {\rm K}$. The temperature $T$ in eq.\,(\ref{T1.4}) as a function of the concentration of anesthetics in the fluid phase, $x_A^f$, is shown as the solid curve in Fig.\,\ref{Figure1}.  It is the phase boundary of the gel-fluid coexistence regime, where the concentration of anesthetics in the gel phase is $x_A^g=0$ by definition. Here, $x_A$ is the fraction of anesthetics in the total membrane. The dashed line in Fig.\,\ref{Figure1} represents the freezing point depression law of eq.\,\ref{T1.5}). This approximation is reasonable for molar fractions of anesthetics as high as approximately $x_A=0.2$.  It fails at higher concentrations where shifts of $T_m$ are considerably larger than predicted by the linear approximation. The effective anesthetic dose in the membrane for tadpoles corresponds to $x_A^f\approx$ 0.026 \cite{Heimburg2007c}. Thus, the freezing point depression law is a good approximation for medically relevant concentrations. The grey shaded region in Fig.\,\ref{Figure1} corresponds to the regime where phase separation between gel and fluid phase takes place.

The fraction of fluid membrane, $x_f$, and of gel membrane, $x_g$, is given by the lever rule \cite{Lee1977}:
\begin{equation}
\label{T1.6}
x_f=\frac{x_A-x_A^g}{x_A^f-x_A^g}  \qquad  ;  \qquad x_g=\frac{x_A^f-x_A}{x_A^f-x_A^g} \;.
\end{equation}
with $x_f+x_g=1$. Since $x_A^g=0$ by definition, we obtain
\begin{equation}
\label{T1.7}
x_f=\frac{x_A}{x_A^f}  \qquad  ;  \qquad x_g=\frac{x_A^f-x_A}{x_A^f} \;.
\end{equation}
The application of the lever rule is demonstrated in Fig.\,\ref{Figure1} for the example of $x_A=0.2$ and $T=300$\,K (filled circle) yielding a fluid fraction of $x_A^f=0.47$. Using  eq.\,(\ref{T1.7}), we find that $x_f=0.426$ and $x_g=0.574$.  Fig.\,\ref{Figure1} also indicates that the concentration of anesthetics in the fluid membrane is a function of temperature. The lower the temperature, the smaller the fluid fraction and therefore the higher the concentration of anesthetics $x_A^f$ in the fluid phase.

\begin{figure*}[bht!]
    \centering
	\includegraphics[width=16cm]{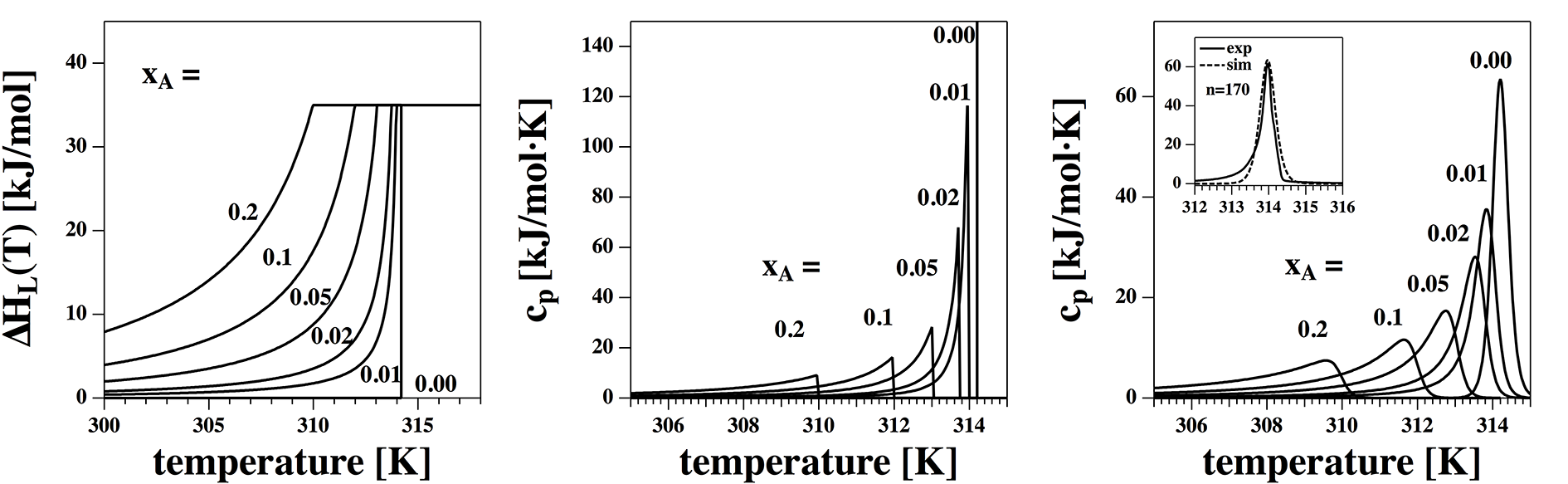}
	\parbox[c]{17cm}{ \caption{\textit{Left:\ The enthalpy of DPPC calculated using eq.\,(\ref{T1.8}) for six different mol fractions $x_A$ of a general anesthetic. Center: The derivative of the enthalpies in the right panel yields the corresponding heat capacity profiles. Right: Broadened $c_p$ profiles. The insert shows the experimental heat capacity profile of DMPC LUV (downscan), and a van\,'t Hoff profile using a cooperative unit size of $n=170$ generated using eq. (\ref{T1.9}). The latter curve was used to convolute the $c_p$-profiles shown in the center panel. }
	\label{Figure2}}}
\end{figure*}

This allows us to calculate the melting profile. The enthalpy per mole of lipid is given by
\begin{equation}
\label{T1.8}
\Delta H_{L} (T) = x_f\cdot x_L^f\cdot \Delta H_{L,0} \cdot\frac{1}{x_L}
\end{equation}
$x_L=1-x_A$ is the fraction of lipid in the total membrane and is used for normalization to 1 mol of lipid. The heat capacity is given by the derivative, $\Delta c_p^L=(d\Delta H/dT)_p$. Enthalpy and heat capacity are given for various values of the anesthetics concentration in Fig.\,\ref{Figure2} (left and center).  The heat capacity profiles are asymmetrically broadened towards lower temperatures. This is a consequence of the increase of anesthetic concentration in the remaining fluid regions of the membrane upon cooling.

\subsubsection*{Convolution with the natural width of the transition profile}
In the above formalism the heat capacity of a membrane in the absence of anesthetics is represented by a $\delta$-function. This is a consequence of the assumption of macroscopic phase separation in ideal solution theory. However, experimental profiles typically display a finite transition width. The result for DPPC-LUV it is shown in Fig.\,\ref{Figure2} (right, insert).  If phases do not separate macroscopically but rather into domains with a cooperative unit size of $n$, the temperature-dependent enthalpy of the pure membrane is given as \cite{Heimburg2007a}
\begin{eqnarray}
\label{T1.9}
\Delta H_{L}^{peak} (T) &=& \frac{K}{1+K}\cdot \Delta H_{L,0}\quad;\\
K=K(T,T_m)&=&\exp\left[-\frac{n \Delta H_{L,0} }{R}\left(\frac{1}{T}-\frac{1}{T_m}\right)\right]\nonumber
\end{eqnarray}
The temperature derivative is the heat capacity $c_p^{peak}$:
\begin{equation}
\label{T1.10}
\Delta c_p^{peak}(T) = \frac{K(T,T_m)}{[1+K(T,T_m)]^2}\frac{n\Delta H_{L,0}^2}{RT^2} 	\;,
\end{equation}
which has an integrated enthalpy of $\Delta H_{L,0}$. For unilamellar DPPC vesicles is shown in Fig.\,\ref{Figure2} (right, insert). A $\delta$-function peak in ideal solution theory at $T_m=41.1^{\circ}C$ leads to the above peak shape in a calorimetric experiment. The integrated enthalpy of both is $\Delta H_{L,0}$. A cooperative unit size of $n=170$ was found to yield a satisfactory transition width for the down scans of DPPC LUV. This value was used for all convolutions shown in this paper. 

In order to be able to compare the theoretical with experimental heat capacity profiles, we convoluted the theoretical profiles $\Delta c_p^L(\tau)$ with the transition profile of DPPC LUV $\Delta c_p^{peak}$ given by eq.\,(\ref{T1.10}):
\begin{eqnarray}
\label{T1.11}
\Delta c_p^{broad} (T) =\int_{\tau=0}^{+\infty}\Delta c_p^L(\tau)\frac{K(T, \tau)}{(1+K(T, \tau))^2} \frac{n\Delta H_{L,0}}{RT^2}d\tau \;; \nonumber\\
K(T, \tau)=\exp\left(-\frac{n \Delta H_{L,0} }{R}\left(\frac{1}{T}-\frac{1}{\tau}\right)\right) 	\;.
\end{eqnarray}

In Fig.\,\ref{Figure2} (right) we show the theoretical profiles in the center panel convoluted with the function given in eq.\,(\ref{T1.10}) using the above procedure. We also applied this formalism in Figs.\,\ref{Figure2c}, \ref{Figure4}, \ref{Figure5} and \ref{Figure6} (see below).

\subsection*{2. Infinite reservoir size}\label{Theory_2}
In the previous section we assumed that all anesthetics stay in the membrane, i.e., the total number of anesthetic molecules in the membrane is constant. This is the case either if the partition coefficient of the drug is very high or if the amount of aqueous medium is very small. 

\begin{figure}[htb!]
\centering
	\includegraphics[width=8.5cm]{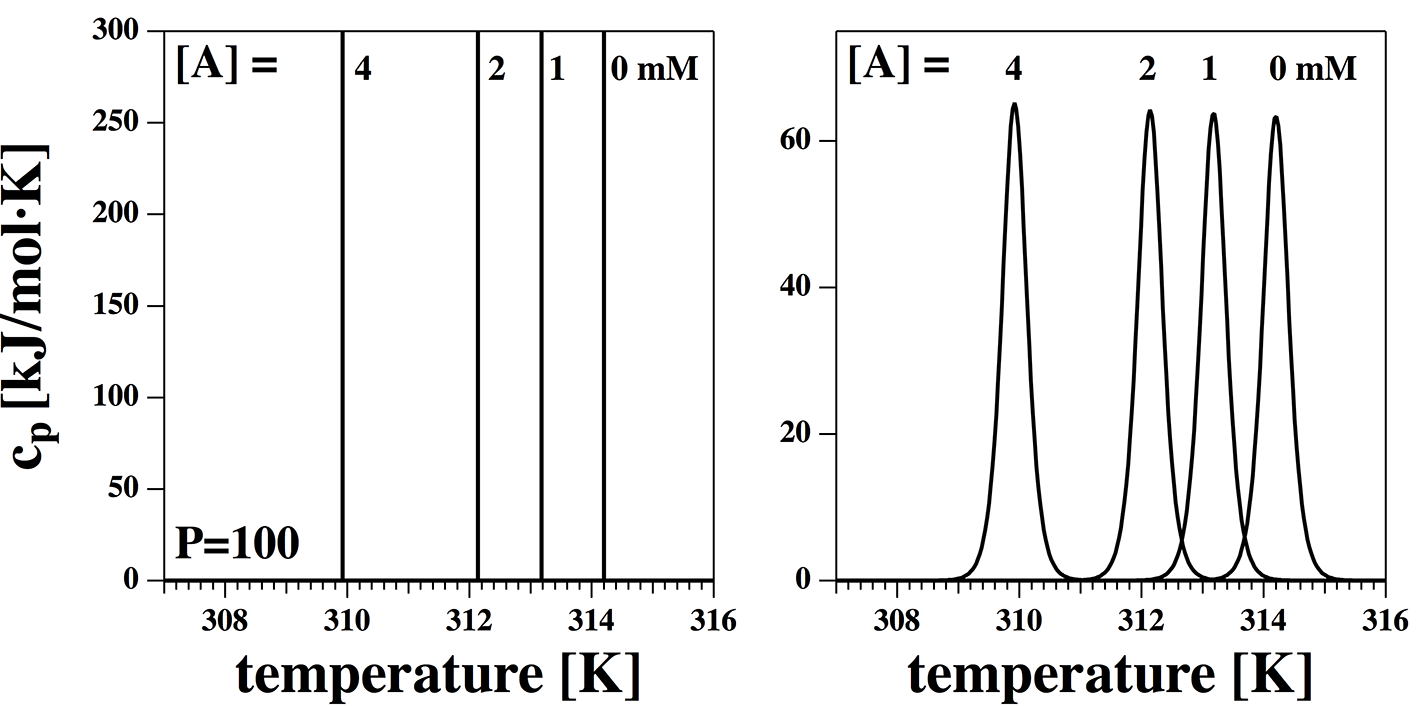}
	\parbox[c]{8.5cm}{ \caption{\textit{Calorimetric profiles in an infinite reservoir for  four different aqueous concentrations of anesthetic, $[A]$, assuming a partition coefficient $P=100$. Left: In ideal solution theory, the calorimetric peak is a $\delta$-function which is shifted in the presence of anesthetics. Right: When convoluted with the profile of DPPC LUV  (see eq.\,(\ref{T1.11})) one sees that the peak shape is practically unaffected by the presence of anesthetics.}
	\label{Figure2c}}}
\end{figure}

Let us consider an infinite reservoir and a finite partition coefficient. Now, anesthetic molecules can be exchanged between the aqueous medium and the fluid membrane such that the concentration of anesthetics in the fluid membrane is fixed. This is true because an infinite reservoir displays a constant concentration of anesthetics, and the concentration in the fluid membrane is determined by the partition coefficient alone. The total amount of anesthetics in the membrane is proportional to the fraction of fluid phase. In ideal solution theory, the melting peak is a sharp $\delta$-function. In the presence of anesthetics the melting profile will not be asymmetrically broadened but will still shift by the value given in eq.\,(\ref{T1.5}). This situation is shown in Fig.\,\ref{Figure2c}.
\subsection*{3. Finite size reservoir}\label{Theory_3}

The most general case is that of a finite aqueous reservoir and a finite partition coefficient. This is the situation in most calorimetric experiments. When the membrane melts, the volume of the fluid phase changes, and it can absorb more anesthetic molecules. Since the reservoir is finite, the anesthetic concentration there decreases.  Neither the concentration of the anesthetic drug in the fluid membrane nor in the total membrane stay constant.  The considerations of section 1 remain valid, i.e., the ratio of the sizes of fluid and gel phases of the membrane for a given concentration of anesthetics in the fluid phase is given by the lever rule of eq.\,\ref{T1.7}. However, the total amount of anesthetics in the membrane and the concentration in the bulk aqueous medium can vary with changes in the amount of fluid phase. 
\begin{figure}[htb!]
    \centering
	\includegraphics[width=8.5cm]{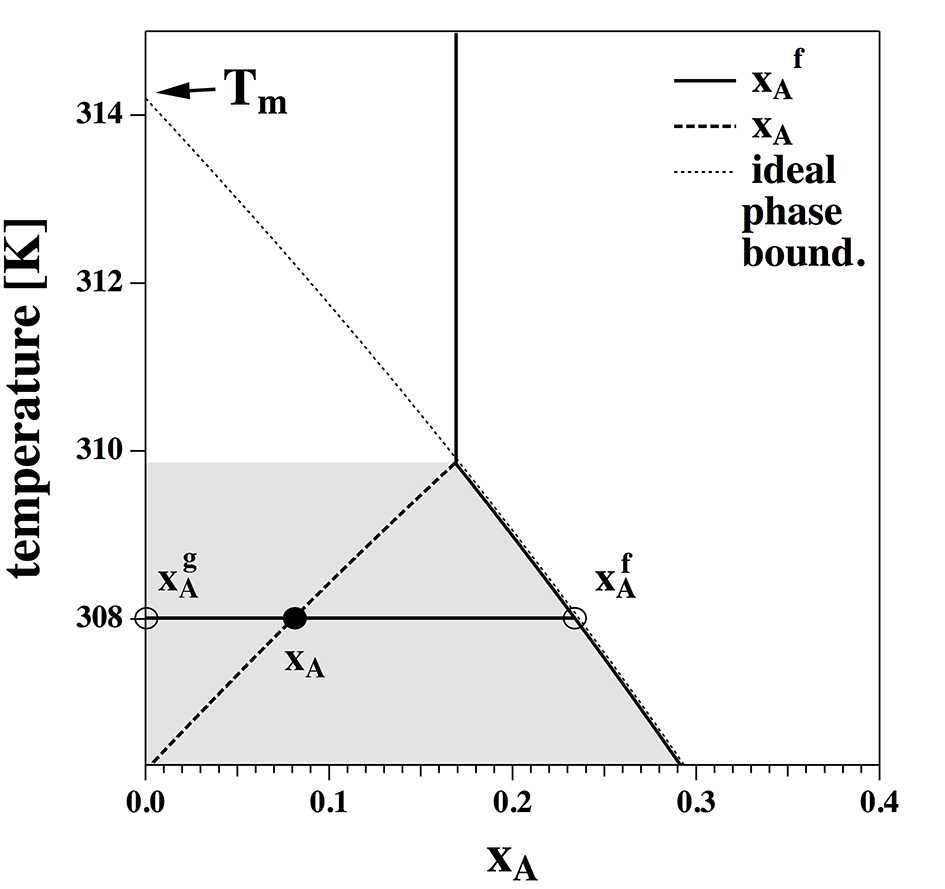}
	\parbox[c]{8.5cm}{ \caption{\textit{Calculation of $x_A$ and $x_A^f$ for case 3. In this example we used 10 mM DPPC, $P=100$, $[A_{tot}]=4$ mM general anesthetic ($m_T^A= 4\, \mu$mole in 1 ml volume).  When the membrane is entirely fluid, i.e., at high temperature, this corresponds to $x_A=x_A^f = 0.17$ (vertical solid line). Upon cooling, one hits the phase boundary at $T=309.9$ K. Below this temperature, $x_A^f$ increases along the phase boundary while the total anesthetic fraction in the membrane, $x_A$, decreases due to the decrease of the fluid membrane fraction,  $x_f$. At T=306.2 K, $x_A=0$ and $x_f=0$. }
	\label{Figure3}}}
\end{figure}

Let us assume a total concentration of anesthetics $[A_{tot}]$ and a total volume of the sample of $V_{tot}$. The total molar quantity of anesthetic, $m_T^A=[A_{tot}]\cdot V_{tot}$, is given by
\begin{equation}
\label{T3.1}
m_T^A=\underbrace{P\cdot[A]\cdot V_{L}^f}_{\mbox{\tiny molar amount in membrane}}+\underbrace{[A]\cdot V_{B}}_{\mbox{\tiny molar amount in buffer}}
\end{equation}
where $[A]$ is the free anesthetic concentration, $P$ is the partition coefficient,  and $V_B$ is the volume of aqueous buffer, $V_L$ the volume of the lipid membrane with $V_{tot}=V_B+V_L$.  The volume of fluid membrane, $V_L^f$, is given by
\begin{equation}
\label{T3.2}
V_L^f=V_L\cdot x_f
\end{equation}
where $x_f$ is the fluid membrane fraction. The total amount of anesthetics, $m_T^A$, is fixed during an experiment and independent of temperature. We further assume that $V_B$, $V_L$, and $P$ are known and fixed. The variables are $[A]$ and $x_f$. For a given $[A]$, the fluid membrane fraction $x_f$ can be determined  using eqs.\,(\ref{T3.1}) and (\ref{T3.2}):
\begin{equation}
\label{T3.2b}
x_f=\frac{m_T^A-[A]\cdot V_B}{P\cdot [A]\cdot V_L}
\end{equation}
The molar fraction of anesthetics in the fluid membrane is given by 
\begin{equation}
\label{T3.3}
x_A^f=P\cdot[A]\cdot v_L^0
\end{equation}
where $v_L^0$ is the molar volume of the fluid lipid membrane (approximately 0.734\,l/mol for DPPC assuming a density of 1 g/cm$^3$). For a given $[A]$, one can calculate the concentration of anesthetics in the fluid membrane (using eq.\,(\ref{T3.2b})), and the fluid fraction $x_f$ (using eq.\,(\ref{T3.2})). When $x_A^f$ is known, one can calculate the corresponding temperature, $T$, of the phase boundary using eq.\,(\ref{T1.4}):
\begin{eqnarray}
\label{T3.4}
T&=&\left[\frac{1}{T_{m,L}}-\frac{R}{\Delta H_{L,0}}\ln(1-x_A^f)\right]^{-1}\nonumber\\
&=&\left[\frac{1}{T_{m,L}}-\frac{R}{\Delta H_{L,0}}\ln(1-P\cdot [A] \cdot v_L^0\cdot )\right]^{-1}
\end{eqnarray}
According to eq. (\ref{T1.7}), the total anesthetic fraction $x_A$ in the membrane can be derived from the lever rule:
\begin{equation}
\label{T3.5}
x_A= x_f\cdot x_A^f	,
\end{equation}
which is a function of $[A]$.  If we regard the total anesthetic concentration $[A_{tot}]$, the total lipid concentration (or total lipid volume $V_L$) and the partition coefficient $P$ of the anesthetic in the fluid membrane as fixed input parameters, we can determine phase boundary temperature, the fluid membrane fraction $x_f$, the fluid membrane fraction of the anesthetic $x_A^f$ and the anesthetic concentration $x_A$ for any given value of [A]. The problem is completely determined. In Fig.\,\ref{Figure3} we demonstrate the change of both $x_A$ and $x_A^f$ as a function of temperature for the case of $P=100$, 10 mM lipids and a total anesthetics concentration of 4 mM. Upon decreasing temperature, $x_A= x_A^f$ stays constant until the upper phase boundary is reached. Upon further cooling, $x_A^f$ increases while the total fraction of anesthetics in the decreases until it becomes zero. This constitutes a lower end of the melting profile.

\begin{figure*}[htb!]
    \centering
	\includegraphics[width=12cm]{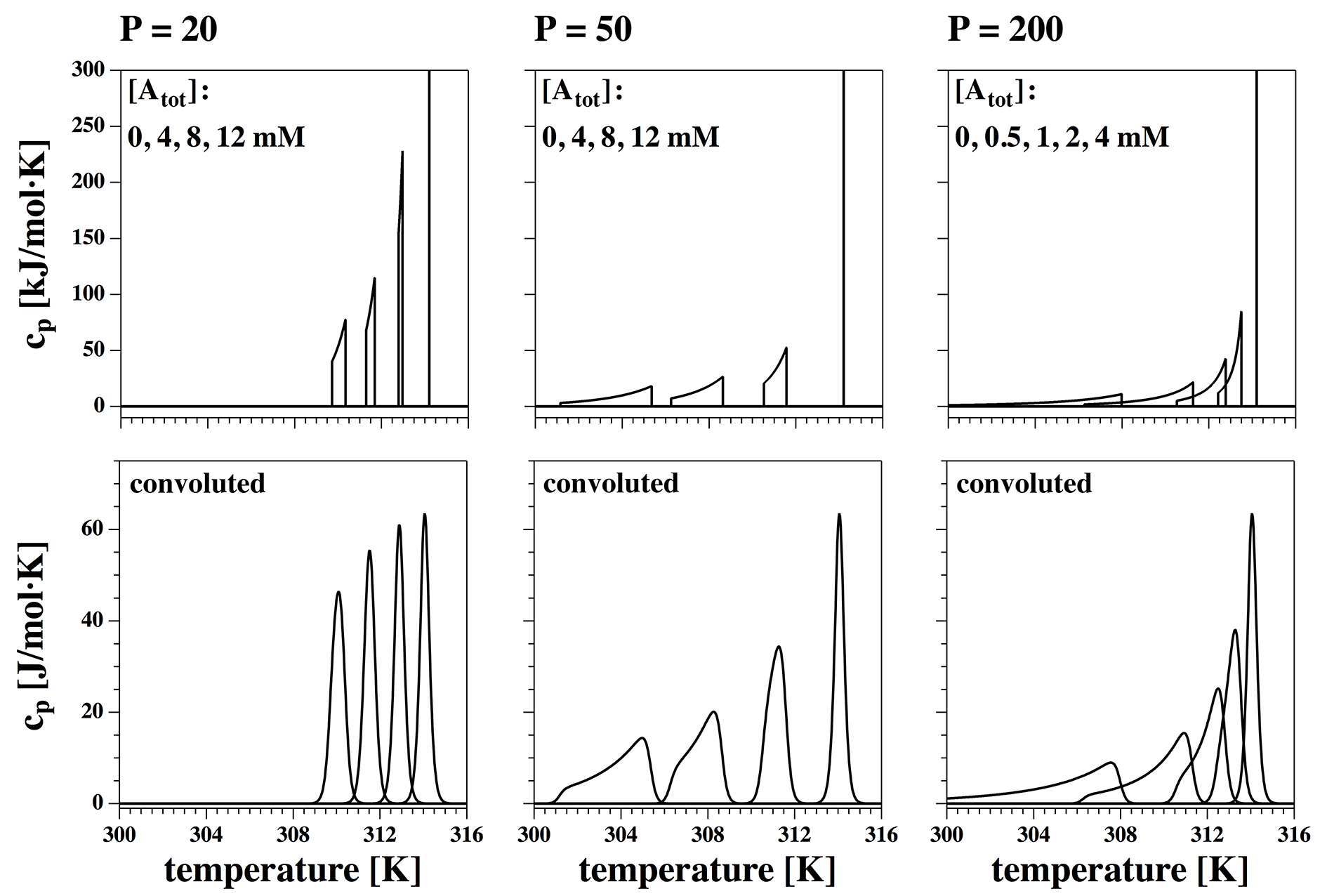}
	\parbox[c]{17cm}{ \caption{\textit{Calculation of heat capacity profiles for case 3 using eq.\,(\ref{T3.6}) for three different partition coefficients and various anesthetics concentrations. The DPPC lipid concentration was assumed to be 10\,mM. The top row contains the theoretical calculations and the bottom row the broadened profiles for DPPC LUV following the convolution procedure in eq.\,(\ref{T1.11}). Left: P=20. Center: P=50. Right: P=200.  The shape of the $c_p$ profile depends sensitively on the partition coefficient.}
	\label{Figure4}}}
\end{figure*}

The melting profile of the membrane can now be determined in analogy to eq.\,(\ref{T1.8}):
\begin{equation}
\label{T3.6}
\Delta H_{L} (T) = x_f\cdot (1-x_A^f) \cdot \Delta H_{L,0} \cdot\frac{1}{(1-x_A)}
\end{equation}
where $x_A$ is the fraction of anesthetic in the total membrane, which is a function of temperature. The final term here is needed for normalization to 1 mol of lipid. The heat capacity is the temperature derivative of this function. Fig.\,\ref{Figure4} shows the heat capacity profiles for three different partition coefficients $P$ (= 20, 50 and 200) using various  total anesthetics concentrations, $A_{tot}$. The top panels show the results using the above formalism. One can clearly see the upper and lower limits of the heat capacity anomaly. The lower limit of the melting profile did not exist in case\,1 (Fig.\,\ref{Figure2}). The bottom panels of Fig.\,\ref{Figure4} display the corresponding convolutions with the profile of unilamellar vesicles. 


\section*{Experimental Results}
In the following we compare calorimetric profiles of DPPC LUV obtained in the absence and presence of the general anesthetic octanol, the barbiturate pentobarbital, and the two local anesthetics lidocaine and bupivacaine with theoretical calculations as described above. We use experimental DSC down-scans obtained with a scan rate of 5 deg/hr.

In the calorimetric experiment, the total aqueous volume, the lipid concentration, and the the total anesthetic concentration are fixed. We assumed a lipid membrane density of $1$\,g/cm$^3$ in order to calculate the total lipid volume. The molar volume of DPPC is then $v_L^0=0.734$\,l/mol (MW=734 g/mol). The lipid volume of 1 ml of a 10\,mM dispersion is $V_L=7.34\,\mu$l and the volume of the aqueous buffer is $V_B=992.66\,\mu$l. For a 1\,mM solution, the total molar quantity of anesthetics is $m_T^A=1\,\mu$mol. The only unknown parameter in eqs.\,\ref{T3.1}--\ref{T3.6} is the partition coefficient, $P$.
\begin{figure}[hb!]
    \centering
	\includegraphics[width=8.5cm]{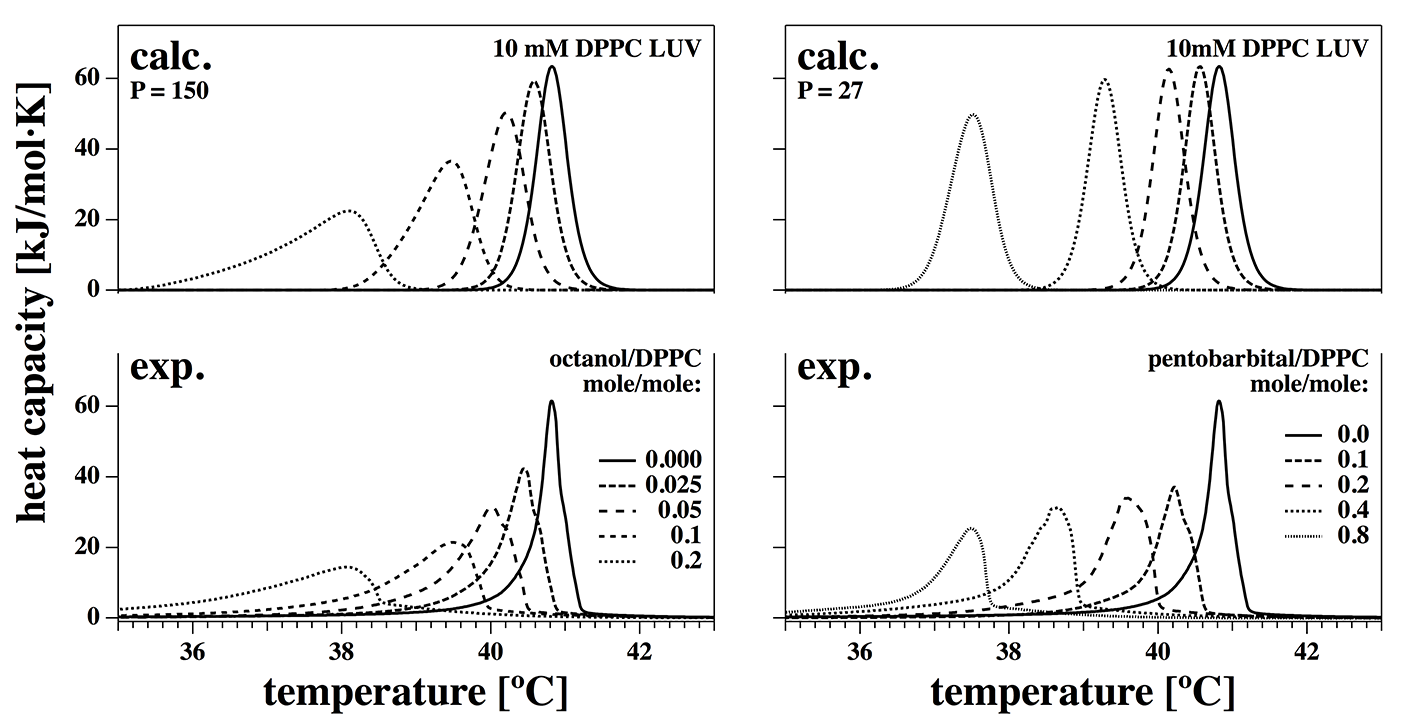}
	\parbox[c]{8.5cm}{ \caption{\textit{Heat capacity for DPPC LUV in the presence of two general anesthetics (octanol and pentobarbital). Experimental heat capacity profiles for 10mM DPPC LUV in the presence of five concentrations of octanol (bottom left) and four concentrations of the barbiturate pentobarbital (bottom right). The upper panels indicate the theoretical results at the same concentrations and the partition coefficient best suited to describe the experimental results. The extracted partition coefficients in the DPPC membrane are P=150 for octanol and P=27 for pentobarbital.}
	\label{Figure5}}}
\end{figure}

Fig.\,\ref{Figure5} (left, bottom) shows experimental heat capacity profiles for 10mM DPPC LUV dispersions in the presence of the general anesthetic octanol. The total octanol concentrations $[A_{tot}]$ (in buffer and membrane combined) were 0\,mM, 0.25\,mM, 0.5\,mM, 1\,mM and 2\,mM (corresponding to molar octanol/lipid ratios of 0, 0.025, 0.05, 0.1 and 0.2). Fig.\,\ref{Figure5} (left, top) shows the corresponding simulations using a partition coefficient of $P=150$. This value is similar to the literature value of $P=150$ in erythrocytes \cite{Seeman1971, Seeman1972}. One sees that the decay of the simulated profiles is somewhat less pronounced than that of the experimental profiles, which may be due in part to the difference between experiment and the theoretical $c_p$ profile of the pure DPPC LUV in the absence of anesthetics used for the convolution of the theoretical results. Qualitatively similar results have been reported for halothane in DPPC vesicles \cite{Mountcastle1978}.  Fig.\,\ref{Figure5} (right, bottom) shows the experimental heat capacity profiles for 10mM DPPC LUV in the presence of the barbiturate pentobarbital.  $[A_{tot}]$ concentrations were 0\,mM, 1\,mM, 2\,mM, 4\,mM and 8\,mM (corresponding to molar pentobarbital/lipid ratios of 0, 0.1, 0.2, 0.4 and 0.8). Fig.\,\ref{Figure5} (right, top) show simulations using a partition coefficient of $P=27$. 
\begin{figure}[hb!]
    \centering
	\includegraphics[width=8.5cm]{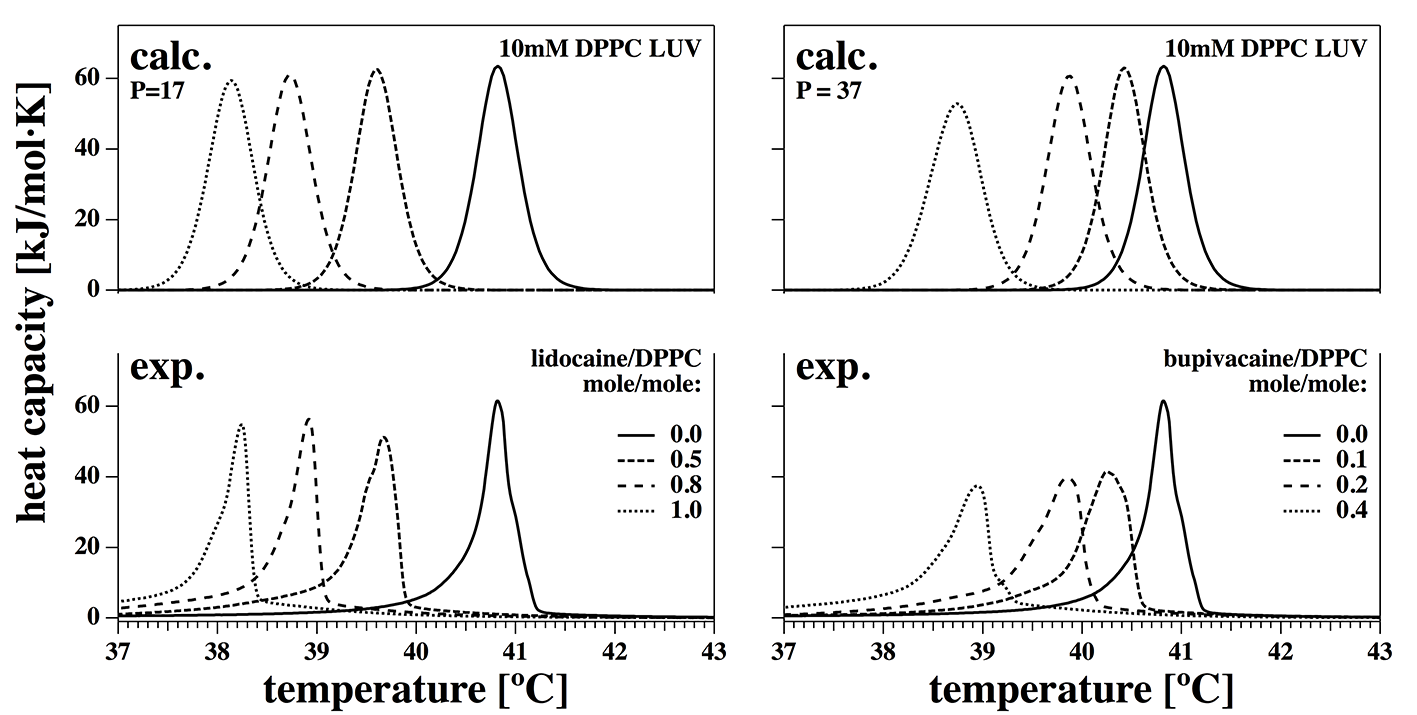}
	\parbox[c]{8.5cm}{ \caption{\textit{Heat capacity of DPPC LUV in the presence of two local anesthetics (lidocaine and bupivacaine): Experimental heat capacity profiles of 10mM DPPC LUV in the presence of four concentrations of lidocaine (bottom left) and four concentrations of bupivacaine (bottom right). The panels on the top indicate the theoretical results at the same concentrations and the partition coefficient best suited to describe the experimental results.  The extracted partition coefficients in the DPPC membrane are P=17 for lidocaine and P=37 for bupivacaine.}
	\label{Figure6}}}
\end{figure}

Fig.\,\ref{Figure6} shows experimental heat capacity profiles for 10mM DPPC LUV dispersions in the presence of the two local anesthetics lidocaine (left) and bupivacaine (right).  For lidocaine (Fig.\,\ref{Figure6}, left, bottom) we used $[A_{tot}]$ concentrations of 0 mM, 5 mM, 8 mM, and 10 mM (corresponding to molar lidocaine/lipid ratios of 0, 0.5, 0.8, and 1.0).  A partition coefficient of 17 yielded the best description of the data. Fig.\,\ref{Figure6} (left, top). 
Our calorimetric results for lidocaine agree well with the calculated values and resemble those reported by \cite{Ueda1994} for 3mM DPPC LUV.
For bupivacaine, Fig.\,\ref{Figure6} (right, bottom), we used $[A_{tot}]$ concentrations of 0\,mM, 1\,mM, 2\,mM, and 4\,mM (corresponding to molar bupivacaine/lipid ratios of 0, 0.1, 0.2, and 0.4).  A partition coefficient of 37 yielded a good description of the data (Fig.\,\ref{Figure6}, right, top).   

\subsection*{Comparison of octanol/water partition coefficients with calorimetric values}

\begin{figure}[hb!]
    \centering
	\includegraphics[width=8.0cm]{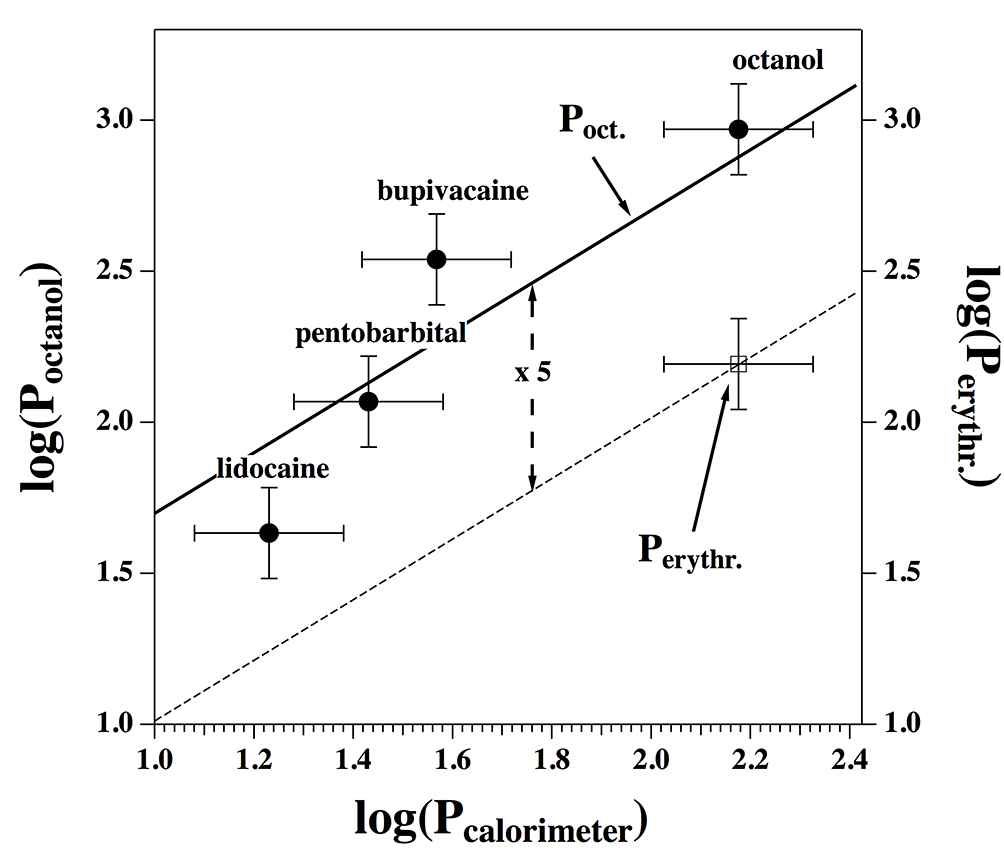}
	\parbox[c]{8.5cm}{ \caption{\textit{Octanol-water partition coefficients plotted versus the partition coefficient determined by calorimetry (filled circles). The solid line assumes that the two partition coefficients are proportional. The erythrocyte/water partition coefficient for octanol is given for comparison (open square, dashed line). The octanol/water partition coefficients differ by a factor 5 from the erythrocyte/water coefficients (dashed line). The symbols correspond to bold and underlined values in Table \ref{Table1}}.
		\label{Figure9}}}
\end{figure}

Partition coefficients from the literature and those obtained from our calorimetric experiments are compared in Table \ref{Table1} and in Fig.\,\ref{Figure9}. The partition coefficients $P$ are defined as the ratio of the concentration of either the dissolved charged or uncharged substances in the organic phase and the concentration in the aqueous phase. Thus, for drugs with a pK$_A$ close to experimental conditions we display partition coefficients $P_c$ for the charged form and $P_u$ for the uncharged form. The distribution coefficient $Q$ corresponds to the partition coefficient for both forms combined.

Partition and distribution coefficients vary with temperature. We compared the results from DSC with octanol/water partition coefficients obtained under similar conditions (i.e., between room temperature and 25 $^\circ$C and at a pH close to the chosen value of 7). These values are shown in table \ref{Table1} as bold underlined numbers. Assuming a proportional relation between the two quantities, the calorimetric values are by factor of 5 smaller than the octanol partition coefficients, which is in very good agreement agreement with the values by \cite{Seeman1971, Seeman1972} who  found a factor 5 difference in the octanol and the erythrocyte. The value for the partitioning of octanol in erythrocyte membranes is given in Fig.\,\ref{Figure9} for comparison. The spread of literature values given in Table \ref{Table1} is considerable and is of least of the order of a factor of two. This is tentatively taken into account by the error bars in Fig.\,\ref{Figure9} that indicate the range of a factor two (end to end of the error bars). Some of the uncertainty may be due to the pH dependence of pentobarbital, lidocaine and bupivacaine that have pK$_A$ values of about 8.2. The pH in our experiments was 7.0.  At this pH, pentobarbital is in its uncharged form (membrane-soluble) while lidocaine and bupivacaine are in the charged (water-soluble) form. Thus, the experimental situation is seemingly far away from the pK$_A$. However, the fact that charged and uncharged forms have different partition coefficients immediately implies that the pK$_A$ of these substances in the membrane must be different from that in solution. In the following scheme we consider an anesthetic drug that is uncharged in its protonated form.  $A^-$ denotes a charged anesthetic molecule, $H^+$ are protons, and $M$ is the membrane. 
\begin{table}[ht]
	\tiny
	\centering
	\caption{\textit{Partition coefficients of octanol, pentobarbital, lidocaine and bupivacaine. The distribution coefficient, $Q$, is the ratio of the concentration of molecules in the organic phase to the concentration  of charged and uncharged molecules in aqueous phase).  $P_u$ and $P_c$ are the partition coefficients of the uncharged and charged form, respectively. RT is room temperature. The bold, underlined values are used in Fig.\,\ref{Figure9}}}
	\vspace*{0.5cm}
	
	\label{Table1}
	\begin{tabular}{cccccccccc}
	\toprule
	substance	&	$Q$	&	$P_u$	&	$P_c$	&	solvents	&	pH	&	T	&	pK$_A$	&	Ref.\\
	\midrule
	octanol	&	1000	&	&	&	n-oct.:water	&	&				&	&	\cite{Hansch1995}\\ \noalign{\smallskip}
			&	1410	&	&	&	n-oct.:water	&	&	25$\,^{\circ}$C	&	&	\cite{Collander1951}\\ \noalign{\smallskip}
			&	\underline{\textbf{933}}&		&	&	n-oct.:water	&	&	22$\,^{\circ}$C	&	&	\cite{Konemann1979}\\ \noalign{\smallskip}
			&	691	&	&	&	n-oct.:water	&	&	&	&	\cite{Leo1969}\\ \noalign{\smallskip}
			&	1810	&	&	&	n-oct.:water	&	&	45$\,^{\circ}$C		&		&	\cite{Rowe1998}\\\noalign{\smallskip}
			&	\underline{\textbf{156.04}}	&	&	&	erythr.:buffer	&	7.0	&	RT	&		&	\cite{Seeman1971}\\\noalign{\smallskip}
			&	151.8	&	&	&	erythr.:buffer	&	&	&	&	\cite{Seeman1972}\\\noalign{\smallskip}
			&	\underline{\textbf{150 (cal.)}}	&	&	&	DPPC:buffer	& 7.0 	&	&	&	this work\\
	\midrule
	pentobarbital	&	&	\underline{\textbf{117}}	&	&	n-oct.:water	&	&	25$\,^{\circ}$C	&		&	\cite{Wong1988b}\\ \noalign{\smallskip}
			&	53.7	&	135	&	&	n-oct.:water	&	7.4	&	&	&	\cite{Yih1977}\\ \noalign{\smallskip}
			&	9.6	&	&	&	erythr.:buffer	&	7.4 &	23$\,^{\circ}$C	&	&	\cite{Seeman1972}\\ \noalign{\smallskip}
			&	8.5	&	&	&	erythr.:buffer	&	7.4 &	23$\,^{\circ}$C	&	&	\cite{Roth1972}\\ \noalign{\smallskip}
			&	&		&	&		&	&	25$\,^{\circ}$C	&	8.17  &	\cite{Ishihama2002}\\ \noalign{\smallskip}
			&	&		&	&		&	&	37$\,^{\circ}$C	&	7.95  &	\cite{Ballard1961}\\ \noalign{\smallskip}
			&	\underline{\textbf{27 (cal.)}	}&	&	&	DPPC:buffer	& 7.0 	&	&	&	this work\\

	\midrule
	lidocaine	&	\underline{\textbf{43.0}}		&	304	&	0.060	&	n-oct.:buffer	&	7.4	&	25$\,^{\circ}$C	&	8.19	&	\cite{Strichartz1990}\\ \noalign{\smallskip}
			&	47.9	&	&	&	n-oct.:buffer	&	7.4	&	RT&		&	\cite{Nakazano1991}\\
			&	110.0	&	366	&	0.085	&	n-oct.:buffer	&	7.4	&	36$\,^{\circ}$C	&	7.77 	&	\cite{Strichartz1990}\\ \noalign{\smallskip}
			&	&	245	&	&	n-oct.:buffer	&	9.86	&	23$\,^{\circ}$C	&	&	\cite{Broughton1984}\\ \noalign{\smallskip}
			&			&	&				&	&	&	25$\,^{\circ}$C	&	7.86	&	\cite{Truant1959}\\	\noalign{\smallskip}
			&	\underline{\textbf{17 (cal.)}	}&	&	&	DPPC:buffer	& 7.0 	&	&	&	this work\\
	\midrule
	bupivacaine	&	\underline{\textbf{346.0}}	&	2565		&	1.5	&	n-oct.:buffer	&	7.4	&	25$\,^{\circ}$C	&	8.21	&	\cite{Strichartz1990}\\ \noalign{\smallskip}
			&	560.0		&	3420		&	2.0	&	n-oct.:buffer	&	7.4	&	36$\,^{\circ}$C	&	8.10	&	\cite{Strichartz1990}\\ \noalign{\smallskip}
			&	\underline{\textbf{37 (cal.)}	}	&			&		&	DPPC:buffer	& 	7.0 	&				&		&	this work\\
	\bottomrule
	\end{tabular}
\end{table}

\begin{eqnarray}
\label{D1.1}
A^- +H^+ + M  & \stackrel{\Delta\mbox{\small G}_1, \mbox{\small K}_1}{ \leftrightharpoonsfill{30pt} }  & AH + M \nonumber \\
&  & \nonumber \\
\Delta\mbox{\small G}_2,\mbox{\small K}_2\upharpoonleft \downharpoonright \hspace{1cm}&& \hspace{0.5cm}\upharpoonleft \downharpoonright \mbox{\small K}_3, \Delta\mbox{\small G}_3,\\
&  & \nonumber \\
AM^- + H^+  & \stackrel[\Delta\mbox{\small G}_4, \mbox{\small K}_4]{}{ \leftrightharpoonsfill{30pt}}  & \hspace{0.5cm} AHM \nonumber 
\end{eqnarray}
The constants $K_1$ and $K_4$ denote the protonation equilibria in solution and in the membrane with associated free energy changes $\Delta G_1$ and $\Delta G_4$, respectively. The constants $K_2$ and $K_3$ denote the association equilibria of the charged and the uncharged form of the drug to the membrane with the associated free energy changes $\Delta G_2$ and $\Delta G_3$. The free energy is a function of state. Therefore,
\begin{equation}
\label{D1.2}\Delta G_4=\Delta G_1+\Delta G_3-\Delta G_2\qquad;\qquad K_4=\frac{K_1\cdot K_3}{K_2}
\end{equation}
If the association constant of the anesthetic is different for the charged and the uncharged form, the pK$_A$ in the membrane must necessarily be different from that found in solution. For this reason it is not trivial to determine how much anesthetics will associate to the membrane close to the pK$_A$. 

\subsection*{Pressure dependence}
A theory of anesthesia based on lipid phase transitions has the advantage that it implicitly contains an explanation for the pres\-sure-reversal of anesthesia \cite{Heimburg2007c, Heimburg2008}.   Pressure shifts the transitions of lipid membranes because it alters their specific volume upon melting \cite{Heimburg2007a}:
\begin{equation}
\label{R1.1}
\Delta T_m=\frac{\Delta p \Delta V}{\Delta H}T_m
\end{equation}
Here, $\Delta p$ is the change in hydrostatic pressure and $\Delta V$ is the excess volume of the lipid transition. The pressure dependence of lipid phase transitions has been studied in detail and is quantitatively understood \cite{Ebel2001}. Since anesthetics lower transition temperatures and hydrostatic pressure increases them, one expects a reversal of the effect of anesthetics. The critical pressure for the reversal of tadpole anesthesia was calculated to be about 25 bars \cite{Heimburg2007c} which is of an order similar to that found in experiments \cite{Johnson1950, Halsey1975}.

\begin{figure*}[htb!]
    \centering
	\includegraphics[width=16cm]{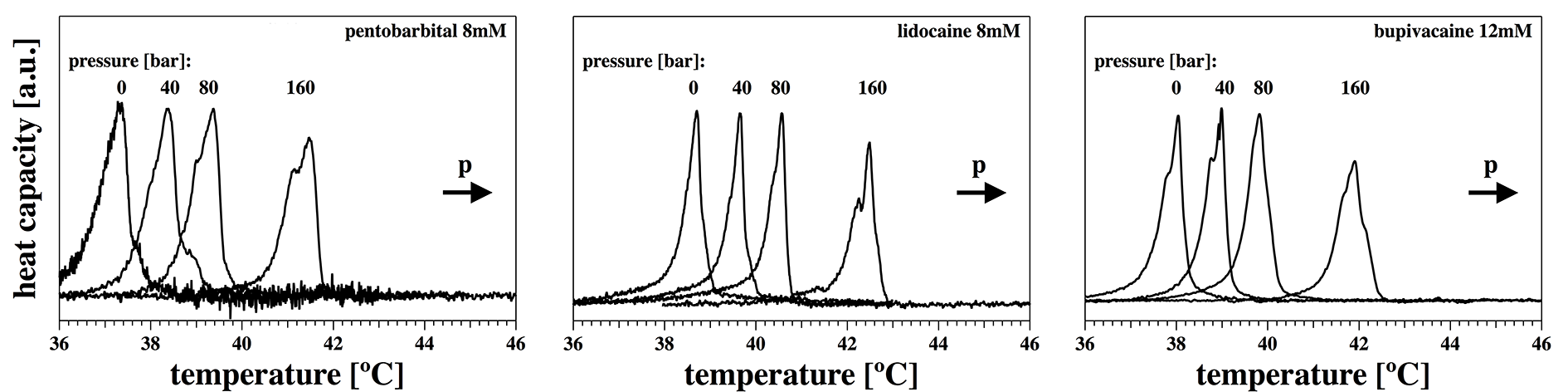}
	\parbox[c]{17cm}{ \caption{\textit{Pressure dependence of membranes in the presence of anesthetics at 0, 40, 80 and 160 bars excess hydrostatic pressure. Left: 10mM DPPC LUV in the presence of a total concentration $[A_{tot}] =$ of 8\,mM pentobarbital. Center: Left: 10mM DPPC LUV in the presence of $[A_{tot}] =$ 8\,mM lidocaine. Right: 10mM DPPC LUV in the presence of $[A_{tot}] =$12\,mM bupivacaine. The shape of the transition profile remains unaltered. The magnitude of the shift is the same as in the absence of anesthetics \cite{Ebel2001}. }
		\label{Figure8}}}
\end{figure*}

As shown in Fig.\,\ref{Figure8}, we have applied hydrostatic pressure to lipid membranes in the presence of both general and local anesthetics. The average shift of the DPPC transition maximum in the presence of three anesthetics (pentobarbital, lidocaine and bupivacaine) in Fig.\,\ref{Figure8} is 0.02441 deg/bar (equivalently, 1 deg/40.95 bar).  This is practically identical to the numbers obtained in the absence of anesthetics as reported in \cite{Ebel2001}, e.g., 0.02448/bar for DPPC. This implies that the pressure dependence of lipid membranes is unaltered by anesthetics. Thus, the solubility of the drugs in liquid and solid phases is  unaltered by hydrostatic pressure within experimental accuracy. 

\section*{Discussion}
General and local anesthetics are generally considered being different classes of drugs. It is further acknowledged that general anesthesia is not well understood. However, it is well known that general anesthetics shift phase transition and generally render membranes more fluid by an mechanism that is independent of the nature of the drug \cite{Kharakoz2001, Heimburg2007c, Heimburg2007b}. In contrast, local anesthetics are often assumed to bind specifically to receptors and in particular to sodium channels \cite{Butterworth1990, Fozzard2011}. It has been known for a long time that also local anesthetics lower lipid melting transitions \cite{Papahadjopoulos1975, Winter1991, Ueda1994, Matsuki1996, Hata2000, Lygre2009, Paiva2012}. Kaminoh et al. \cite{Kaminoh1992} expressed the hypothesis that it is not the solubility in the membrane that decides about anesthetic effects but rather the difference in solubility between fluid and gel phase. This is also the working hypothesis in this paper. 

Here, we studied the effect of octanol, pentobarbital, lidocaine and bupivacaine on the melting transition of unilamellar DPPC vesicles. We demonstrated that general and local anesthetics both lower transition temperature in a qualitatively very similar manner. We provided a formalism to describe the results theoretically. We based our theoretical considerations on ideal solution theory and the assumption that both general and local anesthetics dissolve ideally in the fluid membrane but are insoluble in the gel membrane. We distinguished three cases: 1. All anesthetics dissolve in the membrane either due to a very high partition coefficient or a very small volume of the aqueous buffer.  2. Infinite volume of an aqueous buffer with constant anesthetic concentration. 3. The general case describing a finite amount of buffer and small or medium value partition coefficients. Case one leads to large shifts towards lower temperature with an asymmetric broadening of the $c_p$-profiles that reflects the temperature-dependent change in fluid membrane concentration of the anesthetics. Case two leads to a shift of the $c_p$-profile without broadening and change in peak amplitude. Case three displays both shifts and broadening of the profiles but to a lower extent than in case one. Case three is the typical situation in a calorimetric experiment where the aqueous volume is finite. We demonstrated that with this description one can describe the heat capacity profiles of both classes of anesthetics in a satisfactory manner. We further showed that for up to 20 mol\% of anesthetics within the membrane the above treatment is consistent with the freezing-point-depression law by van't Hoff \cite{vantHoff1886} that implies a linear dependence of the melting point on the concentration of the solute in the fluid phase. However, the shift in the transition is largely underestimated by the freezing-point depression law at higher molar fractions of anesthetics. One can extract partition coefficients that reflect the reported partition coefficients in octanol. Freezing-point depression was similarly used as a mean to determine concentrations of solutes in aqueous solution in the original publication of J. H. van 't Hoff in 1886. 

Finally, we demonstrated that hydrostatic pressure leads to a shift of melting peaks towards higher temperatures without a broadening of the $c_p$-profile. This shift is unaffected by the presence of both general and local anesthetics consistent with findings by \cite{Mountcastle1978}. From this it can be concluded that the solubility of anesthetics drugs in the membrane is generally not pressure-dependent in the range investigated here (pressures up to 200 bars). A similar finding was reported for the local anesthetic tetracaine \cite{Winter1991}. In a previous publication \cite{Heimburg2007c} we argued that the pressure reversal of general anesthesia \cite{Johnson1950} can be quantitatively explained when assuming that the shift of the melting transition induced by anesthetics is counteracted by pressure. The results of the present publication imply that this effect should be similarly true for local anesthesia.  It is interesting to note that Halsey and Wardley-Smith found that general anesthesia induced by the local anesthetic procaine was in fact reversed by hydrostatic pressure in tadpoles \cite{Halsey1975}.

One particularly important feature of the above findings is that the thermodynamics of general and local anesthetics is basically the same. Therefore, there exists no reason in a thermodynamics theory of membranes to distinguish the two classes of anesthetics. 

\subsection*{Additivity of general and local anesthesia}
Both the Meyer-Overton correlation and the law of freezing point depression law are generic linear laws that depend only the concentration of the drug in question and are entirely independent of its chemical nature.  Therefore, these laws explicitly contain the prediction of the additivity of the effect of drugs. This effect is well documented for general anesthetics and was discussed by Overton more than 100 years ago \cite{Overton1901}. In the present publication, we have provided strong evidence that the law of freezing point depression law also applies to local anesthetics. This immediately leads to the prediction that the effects of general and local anesthetics are additive, too. From the thermodynamic point of view the effects of general and local anesthetics on membranes are the same.

There are, in fact, numerous publications providing evidence for the additivity of general and local anesthesia. Himes et al.\ \cite{Himes1977} reported that the critical anesthetic dose for the general anesthetic halothane in dogs can be lowered by 50 \% by plasma concentrations above 10 $\mu$g/ml of the local anesthetic lidocaine. Similarly, nitrous oxide anesthesia was enhanced by lidocaine. The intramuscular administration of the local anesthetics lidocaine and bupivacaine increases the hypnotic effect of the general anesthetic midazolam in humans to a degree exactly proportional to the dose of the local anesthetic \cite{Ben-Shlomo2003}. The effect of bupivacaine is larger than that of lidocaine in agreement with its larger partition coefficient in membranes. The hypnotic effect of the general anesthetic halothane is enhanced by the local anesthetics lignocaine and bupivacaine \cite{Ben-Shlomo1997}. Similarly, \cite{Senturk2002} reported that the critical dose of the general anesthetic propofol in humans was significantly lowered by intramuscular administration of both lidocaine and bupivacaine \cite{Senturk2002}. Along the same lines, \cite{Altermatt2012} reported a significantly lowered critical dose of propofol in the presence of lidocaine. Additivity of general and local anesthesia has also been reported for the effect of bupivacaine on general anesthesia by propofol \cite{Agarwal2004}. Similarly, the critical dose of the general anesthetic cyclopropane is lowered by 40\% in the presence of lidocaine \cite{DiFasio1976}. Further, the critical dose of the general anesthetic isoflurane was lowered linearly with the administered dose of lidocaine in cats \cite{Pypendop2005} with a reduction of more than 50\% for 10 $mu$g/ml lidocaine (in agreement with \cite{Himes1977}. The sedative effect of lidocaine was also discussed by \cite{Szmuk2007}.

The linear dependence of general anesthesia on the dose of local anesthetics suggests that both classes of drugs can induce general anesthesia and that they work by similar mechanisms. It should be noted the generic physical laws are based on ideal solubility of drugs in the membrane and are therefore inconsistent with the idea of specific binding to receptors.

\subsection*{The cutoff-effect of long chain alcohols}
In contrast to the Meyer-Overton correlation, the thermodynamic theory presented here relies on the assumption that anesthetics are perfectly miscible in the fluid phase of lipids and perfectly immiscible in the gel phase.  This implies that some molecules that are soluble in membranes are not anesthetics.  For instance, cholesterol is not an anesthetic even though it dissolves in membranes.  This is because it is soluble in the gel (or liquid ordered) phases and has the effect of increasing the temperature of the lipid phase transition \cite{Halstenberg1998}.  Similarly, although other lipids are themselves soluble in a given membrane, they are not generally anesthetics.  Many lipids display transitions in state close to experimental temperature where one of the states is soluble in the solid phase and the other one in the fluid phase. If the secondary membrane component consists of such a lipid, the phase diagram strongly deviates from the idealized eutectic case shown in Fig.\,\ref{Figure1} \cite{Heimburg2007a}. In particular, if the melting point of a secondary lipid component is higher than that of the predominant lipid species, the melting profile is usually shifted towards higher temperatures. Such considerations also apply to lipid-like molecules such as long chain alcohols. Pringle et al.\ \cite{Pringle1981} report that the anesthetic potency of saturated n-alcohols increases up to dodecanol, and that anesthetic action fails at chain length above a value between 12 and 14.  It is interesting to note that this correlates with the melting-temperature of the pure alcohols. The melting temperature of ethanol is -114 $^\circ$C. The melting temperature of octanol is -14.8 $^\circ$C, that of decanol is +6.9 $^\circ$C, dodecanol has 23.9 $^\circ$C, tetradecanol 38.2 $^\circ$C, and hexadecanol has 49.2 $^\circ$C (values taken from \cite{Lide2005}). All alcohols with chain length 14 or longer display transitions above body temperature (37$^\circ$C).  All 1-alcohols that do have general anesthetic effect display transitions well below physiological temperature. Kharakoz \cite{Kharakoz2001} showed that all 1-alcohols in lipid membranes up to decanol nicely follow the freezing-point depression law.  Experiments from our lab show that tetradecanol fails to follow this correlation but rather increases the transition temperature in DPPC vesicles (not shown).  Similarly, \cite{Kaminoh1992} showed that 1-tridecanol and 1-tetradecanol increase the melting temperature of DPPC, while 1-octanol and 1-decanol decrease it. It thus has to be concluded that the cutoff effect of saturated n-alcohols is in agreement with the present theoretical description.

\subsection*{Lipid channels}
Papahadjopoulos et al.\ \cite{Papahadjopoulos1973} showed that membranes are more permeable in the vicinity of phase transitions, and Antonov et al.\ showed the existence of channel-like conduction steps close to these transitions \cite{Antonov1980}.  Formation of these channels, consisting of small pores in the lipid membrane, is more likely in the vicinity of transitions where area fluctuations are known to be large.  Such ``channels'' exist in lipid membranes in the complete absence of proteins and have conduction properties that are virtually indistinguishable from those reported for protein channels, i.e., conductance and channel lifetimes are of the same order and current-voltage relationships display a similar functional form \cite{Laub2012, Blicher2013}. Close to transitions, these lipid channels can be blocked by general anesthetics as a simple consequence of their influence on the phase transition temperature. For instance, we have demonstrated that channels in DOPC/DPPC mixtures can be blocked by the general anesthetic octanol \cite{Blicher2009} in a manner very similar to the reported blocking of Na$^+$-channels and the acetylcholine receptor by octanol (discussed in \cite{Blicher2009}). It is to be expected that local anesthetics have the potential to block lipid channels because they display a comparable influence on the cooperative melting transition. 

\subsection*{Nerves}
Recently, we proposed that electromechanical solitons (localized pulses) can travel in membranes close to phase transitions \cite{Heimburg2005c, Heimburg2007b, Andersen2009, Lautrup2011}. Such transitions exist in biomembranes slightly below physiological temperature \cite{Heimburg2007a}. Therefore, such pulses were proposed to be related to the action potential in nerves. The distance of physiological temperature to the transition maximum is closely related to the free energy necessary to excite such a soliton \cite{Heimburg2007c}. According to the above, both general and local anesthetics change the transition temperature and thus increase the free energy necessary to excite a pulse, resulting in an increase of the stimulation threshold. 


\section*{Conclusions}
We have show here that general and local anesthetics have similar effects on the phase behavior of lipid membranes. This is consistent with a simple freezing point depression law based on the ideal solubility of anesthetic drugs in the fluid phase and a low solubility in the gel phase.  Therefore, from a thermodynamic perspective, there is no reason to distinguish between general and local anesthetics. The present description is consistent with the cutoff effect of long chain alcohols. The effects of both general and local anesthetics are subject to pressure-reversal.\\

{\small \noindent\textbf{Acknowledgments:} 
We thank to Prof. Andrew D. Jackson from the Niels Bohr International Academy for useful discussions and for a critical reading of the manuscript. This work was supported by the Villum foundation (VKR 022130).\\}


\footnotesize{

}
\end{document}